\documentclass[twocolumn,showpacs,preprintnumbers,amsmath,amssymb]{revtex4}

\usepackage{graphicx}
\usepackage{dcolumn}
\usepackage{bm}

\DeclareMathAlphabet{\mathbi}{OML}{cmm}{b}{it} 
\newcommand{\bel}{\begin{equation}\label}
\newcommand{\ee}{\end{equation}}
\newcommand{\beq}{\begin{eqnarray}\label}
\newcommand{\eq}{\end{eqnarray}}
\newcommand{\bc}{\begin{center}}
\newcommand{\ec}{\end{center}}
\newcommand{\bU}{\mathbi{U}}
\newcommand{\bu}{\mathbi{u}}
\newcommand{\bv}{\mathbi{v}}

\newcommand{\bk}{\mathbi{k}}
\newcommand{\bit}{\begin{itemize}}
\newcommand{\eit}{\end{itemize}}
\newcommand{\ben}{\begin{enumerate}}
\newcommand{\een}{\end{enumerate}} 

\newcommand\shalf{\ensuremath{{\scriptstyle\frac{1}{2}}}}
\newcommand{\Ray}{\mathrm{Ra}}
\newcommand{\Pran}{\mathrm{Pr}}
\newcommand{\lp}{\left(}
\newcommand{\rp}{\right)}
\newcommand{\la}{\langle}
\newcommand{\ra}{\rangle}

\begin{document}

\title{Exponentially growing solutions in homogeneous Rayleigh-B\'enard convection} 

\author{E. Calzavarini$^1$, C. R. Doering$^2$, J. D. Gibbon$^3$,
  D. Lohse$^1$, A. Tanabe$^3$ and F. Toschi$^4$} \affiliation{
  $^{1}${Department of Applied Physics, University of Twente, 7500 AE Enschede, The Netherlands}\\
  $^2${Department of Mathematics and Michigan Center for Theoretical
  Physics, University of Michigan, Ann Arbor MI 48109-1043, USA}\\
  $^3${Department of Mathematics, Imperial College London, London SW7 2AZ, UK}\\
  $^4${IAC-CNR, Istituto per le Applicazioni del Calcolo, Viale del
  Policlinico 137, I-00161 Roma, Italy and INFN, Via Paradiso 12,
  I-43100 Ferrara, Italy}\\} \today
\begin{abstract}
  It is shown that homogeneous Rayleigh-B\'{e}nard flow, i.e.,
  Rayleigh-B\'enard turbulence with periodic boundary conditions in
  all directions and a volume forcing of the temperature field by a
  mean gradient, has a family of exact, exponentially growing,
  separable solutions of the full non-linear system of
  equations. These solutions are clearly manifest in numerical
  simulations above a computable critical value of the Rayleigh
  number. In our numerical simulations they are subject to
  secondary numerical noise and resolution dependent instabilities
  that limit their growth to produce 
  statistically steady turbulent transport.
\end{abstract}
\pacs{$47. 27.- i , 47. 27.Te$}

\maketitle

\noindent
Much effort has been expended in recent decades in addressing the
problem of heat transfer in Rayleigh-B\'{e}nard thermal convection
cells \cite{Kad}. An asymptotic high Rayleigh number heat transport
scaling behavior $Nu \sim Ra^{1/2}$ (perhaps with logarithmic
modifications) has been conjectured as the ultimate regime
\cite{RK62,EAS1,GL00,LT1,dc96}.  While current experimental data for
high Rayleigh numbers are controversial
\cite{CCCHCC,GSNS,Somm99,RCCH,CRC}, numerical simulations have not
been very effective in studying this regime because of difficulties in
dealing with the huge number of degrees of freedom engendered when
Rayleigh numbers reach the order of at least $10^{12}$. Recently, some
of us \cite{LT1,LT2} have studied a tri-periodic convective cell, or
homogeneous Rayleigh-B\'{e}nard (HRB) system, in order to bridge such
difficulties, and to investigate the properties of the convective cell
once the effect of boundary layers has been eliminated.  Model systems
of this sort with hyperviscosity were first investigated
computationally by Borue and Orszag \cite{OB}, and later by Celani
{\em et al} (in two spatial dimensions) \cite{CE} with hyperviscosity
and extra large scale dissipation as well.

In this letter we point out some peculiar properties of the HRB
model that are particularly striking in the low Rayleigh number regime, 
the opposite regime from that studied in refs \cite{LT1,LT2}. 
First we display a family of exact, exponentially growing, separable 
solutions of the full non-linear HRB-system. 
We show that these solutions are clearly manifest in direct numerical
simulations in the Rayleigh number regime above a computable critical value. 
Then by way of a careful numerical precision study we show that these 
may be robust and attracting solutions of the full system of 
partial differential equations.

Here we would like to anticipate that recently a cleverly conducted
series of experiments was designed such as to reduce the influence of
top and bottom plates on the physical core of thermal convection
\cite{EXP}.  In these experiments the temperature gradient in the bulk
of the cell is not imposed but rather, as in fixed-flux convection
\cite{FF}, measured as a dependent parameter.  Interestingly, the $Nu
\sim Ra^{1/2}$ and $Re \sim Ra^{1/2}$ scalings observed are consistent
with HRB simulations in \cite{LT1,LT2}.

The system to be studied is described in terms of the following
partial differential equations 
\beq{I1a} \bu_{t} + \bu\cdot\nabla\bu
+\nabla p &=& \nu \Delta\bu +\mbox{{\bf k}} \,\alpha g
\theta\,,\\
\theta_{t} + \bu\cdot\nabla \theta &=& \kappa \Delta \theta +
\frac{\Delta T}{H} u_{z}\,,
\label{I1b} 
\eq 
where $\bu = (u_{x},\,u_{y},\,u_{z})$ is an incompressible velocity
field, ${\nabla \cdot \bu}=0$, $\nu$ and $\kappa$ and $\alpha g$ are
respectively the kinematic viscosity, thermal diffusivity and the
thermal expansion coefficient times the acceleration due to
gravity. These equations are used to describe the evolution of the
velocity field in a triply-periodic cubic volume $[0,H]^{3}$ in the
presence of a temperature field $T({\bm x},t)={\overline T}({\bm
  x})+\theta({\bm x},t)$. The temperature is expressed as a
fluctuation $\theta$ with respect to a mean profile ${\overline
  T}({\bm x})$ that is \textit{imposed} to be equal to the mean
conductive temperature profile in such a Rayleigh-B\'{e}nard cell;
i.e.  linear and of the form ${\overline T}({\bm x}) = -z\Delta T/H +
\shalf\Delta T$.

When nondimensionalizing lengths with $H$, velocities with $\kappa
/H$, and temperatures with $\Delta T$, equations (\ref{I1a}) and
(\ref{I1b}) can be rewritten as \beq{I2a}
\bU_{t}+\bU\cdot\nabla\bU +\nabla P &=& \Pran\,(\Delta\bU +
\mbox{{\bf k}}\,\Ray\,\Theta)\,,\\
\Theta_{t}+\bU\cdot\nabla\Theta &=& \Delta\Theta + w\,,
\label{I2b}
\eq where $\Theta$, $P$, and $\bU = (u,\,v,\,w)$ are the dimensionless
temperature, pressure, and velocity fields, respectively, and $\Pran
\equiv\nu/\kappa$ and $\Ray \equiv \alpha g H^3 \Delta T \lp \nu
\kappa \rp^{-1}$ are the Prandtl and Rayleigh numbers. In this system
periodic boundary conditions are imposed on \textit{all} the dependent
variables on the cube $[0,1]^{3}$. We consider the equations of motion
(\ref{I2a}) and (\ref{I2b}) with spatially mean-zero initial data for
all $\bU$ and $\Theta$ so that solutions subsequently remain spatially
mean-zero at all times.

With these boundary conditions there is a family of nonlinear
``separable'' solutions of (\ref{I2a}) and (\ref{I2b}) where $\bU$,
$\Theta$ and $P$ are functions only of $x$, $y$ and $t$ but not of the
vertical coordinate $z$.  To see this, let $\bv = \mbox{{\bf i}} u(x,y,t) + 
\mbox{{\bf j}} v(x,y,t)$ and $P=q(x,y,t)$.  Then the divergence-free 
velocity $\bv = (u,\,v)$, and $q$, $w$ and $\Theta$ satisfy
\beq{I3a}
\bv_{t} + \bv\cdot\nabla\bv +\nabla q &=& \Pran \Delta\bv\,,\\
w_{t} +\bv\cdot\nabla w &=& \Pran(\Delta w+ \Ray\,\Theta)\,,\label{I3b} \\
\Theta_{t} + \bv\cdot\nabla \Theta &=& \Delta \Theta + w\,.
\label{I3c} 
\eq 
Equation (\ref{I3a}) is the unforced two-dimensional Navier-Stokes equation 
whose solutions decay to zero exponentially in time \cite{DGbk}.  As $\bv$ 
decays away, equations (\ref{I3b}) and (\ref{I3c}) admit \textit{exact} 
solutions of the form
\bel{I4a}
\left(\begin{array}{c}
w(x,y,t)\\
\Theta (x,y,t)
\end{array}\right)
= 
\left(\begin{array}{c}
w_{0}\\
\Theta_{0}
\end{array}\right)
e^{\lambda t}\sin(k_{x}x+k_{y}y +\phi)\,, \ee with growth rate
\beq{I4b} \lambda = -\shalf(\Pran + 1)k^{2} + \shalf\sqrt{(\Pran +
  1)^{2} k^{4} + 4\Pran(\Ray - k^{4})} \eq 
\noindent
and where $\phi$ is an arbitrary phase. Therefore for $Ra$ above a
critical value $Ra_c = k^4 = \lp 2 \pi \rp^{4}$ the exponential
solutions are unbounded; i.e. $\lambda > 0$. Since $\bk =2\pi\lp n_{x},
n_{y}\rp$, where $n_{\{x,y\}}$ are ($\pm$)-integer wave-numbers, we
can re-label the exponent $\lambda$ as $\lambda(n_x, n_y)$.
Degenerate values of $\lambda(n_x, n_y)$ are possible, corresponding
to different combinations of $n_x$, $n_y$.  The number of positive
$\lambda$ grows asymptotically $\sim Ra^{1/2}$. 
These solutions actually transport unlimited heat because
$Nu \sim \left< w \Theta \right> \sim \exp (2\lambda t)$. Such runaway
solutions and their possible instabilities may be actually the
dominant features of the simulations in \cite{OB}, \cite{CE} and 
in \cite{LT1,LT2} where a $Nu \sim Ra^{1/2}$ dependence was observed.
Therefore the relevance of computations on periodic domains as models
for the bulk of systems with essential boundaries may be arguable.  
Indeed, Borue and Orszag \cite{OB} seemed to hint at this issue when they
remarked ``It turns out that in homogeneous convection [the
heat transport, temperature and velocities] are strongly fluctuating and
intermittent in time.  This fact makes reliable measurements of
[these variables] difficult."

One of the purposes of this letter is to show that solutions in
(\ref{I4a}) appear to be attracting and may dominate the dynamics
in numerical integrations of equations (\ref{I1a}) and (\ref{I1b}) on
fully periodic domains at low $Ra$ numbers.  Resolution and precision
difficulties, however, prevent us from drawing firm conclusions about
the physics behind secondary instabilities that
limit the growth of the runaway solutions to produce statistically steady
turbulent transport in simulations at higher $Ra$. 
This remains an open question subject to further, deeper, investigations.
\begin{figure}[h]
\includegraphics[draft=false,scale=0.7]{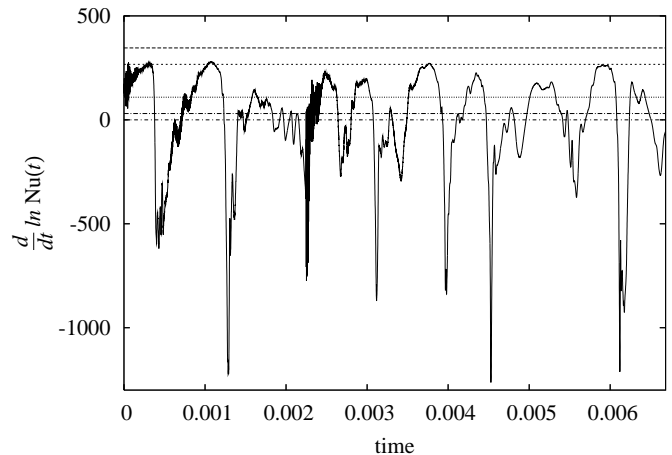}
\caption{Derivative with respect to time of the logarithm of $Nu(t)$,
  in a homogeneous convective system at $Ra = 4.5\times10^4 \sim 30
  \cdot Ra_c$. The DNS is implemented through a LBE algorithm on a
  cubic grid with resolution $240^3$, as in \cite{LT1,LT2}. Horizontal
  lines, from top to bottom, correspond respectively to values
  $2\lambda(0,1)$, $2\lambda(1,1)$, $2\lambda(0,2)$, $2\lambda(0,2)$
  and the zero level.
  The time, here and in the following figures, is dimensionless as for the set of Eq. 
  (\ref{I2a})-(\ref{I2b}),
 i.e., it has been normalized by the thermal diffusion time across the box, $H^2/\kappa$.}
\label{fig1}
\end{figure}
\par\indent
Indeed while unlimited heat transport could be expected because of the
solutions (\ref{I4a}), even at moderately low Rayleigh numbers
simulations of the homogeneous convective system display a statistically
stationary behavior where the growing modes, when they appear, break up
due to rapid destabilization (we limit our investigation to the $Pr=1$
case). Fig.~\ref{fig1} shows the time derivative of the logarithm of
the Nusselt number (i.e. the volume average $Nu(t) = \la w
\Theta\ra_V$) for $Ra \simeq 4.5 \times 10^{4}$. The growth rate of
$Nu(t)$ appears to bounce between some of the admissible exponential
modes, although in these simulations the fastest growing modes, 
$\lambda(0, \pm 1)$ and $\lambda(\pm 1, 0)$, are never reached.  
The numerical results in Fig.~\ref{fig1} were obtained by means of a Lattice
Boltzmann Equation (LBE) algorithm at a relatively high resolution,
$240^3$ (see \cite{LT2} for details).  The value of $Ra$ adopted here
is a little below, but of the same order as, the lower values of $Ra$
for data analyzed in \cite{LT2}.

To better understand the relevant features of the dynamics of the
exponentially growing solutions we have performed a new series 
of direct numerical simulations (DNS) at values of $Ra$ only
slightly above the critical value ($Ra \gtrsim Ra_c$) where
just {\it one} distinct positive value of $\lambda$ exists.
These integrations were performed by means of a fully de-aliased
pseudo-spectral algorithm that allows for more flexibility.  
It allows the adjustment of the time step size that is
implicitly fixed by the spatial grid in the LBE, and gives
clearer control of the scales involved in the dynamics.  
Furthermore, because we are interested here in the
low-$Ra$ regime, it is reasonable to perform 
numerical simulations with lower resolutions (i.e. $32^3$ or $64^3$).
Nevertheless we caution that in case of unlimited exponential growth
{\it any} spatial resolution may be insufficient at some point in time.
\begin{figure}[h]
\includegraphics[draft=false,scale=0.7]{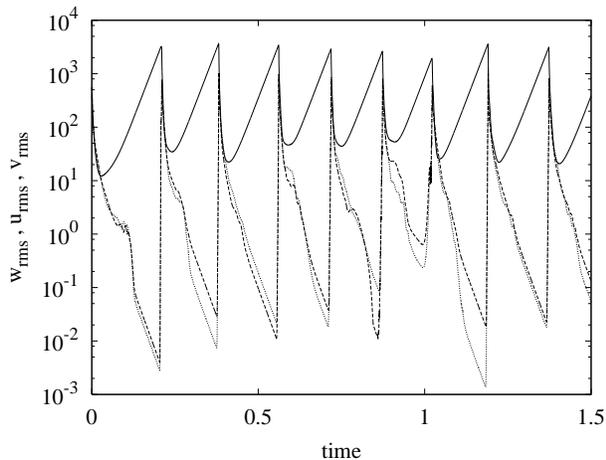}
\caption{Linear-Log plot of the spatial rms-value of the three
  velocity component and the thermal fluctuation in the DNS at
  resolution $32^3$, $w_{rms}$ (solid), $u_{rms}$ (dashed), $v_{rms}$
  (dotted).}
\label{fig2}
\end{figure}
Fig.~\ref{fig2} displays the temporal behavior of the global (spatial)
rms-values of the three velocity components. 
For clarity the temperature $\Theta$
has been omitted from the figure because it is strongly
correlated with $w$, almost coincident with it.
As expected from the unstable analytic solution, $w$
grows at an exponential rate $\propto \exp{(\lambda t)}$ while the horizontal
components $(u,\,v)$ rapidly decrease.
\begin{figure}[h]
\includegraphics[draft=false,scale=0.7]{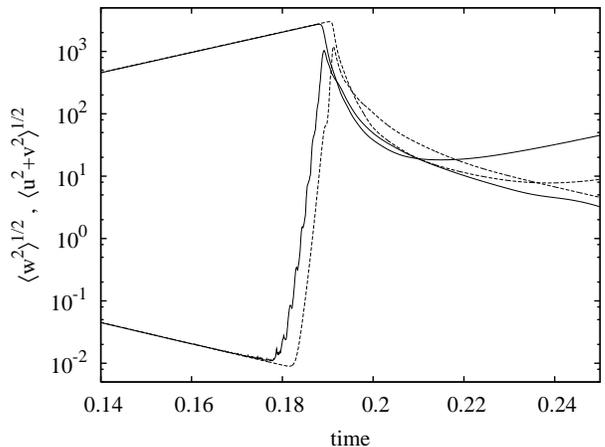}
\caption{A comparison between floating point (solid) and double
  precision (dashed) calculations for the root mean squared velocity
  component $w$, and $\la u^2+ v^2 \ra^{1/2}$ versus time, with
  spatial resolution $32^3$ and 2nd order Adams-Bashforth as time
  marching algorithm.}
\label{fig3}
\end{figure}
The plateaux in Fig.~\ref{fig1} and the linear growth 
(on the lin-log scale) in Fig.~\ref{fig2} indicate the presence of 
exponentially growing solutions, but it is difficult to determine the nature 
of the sudden departures from this state. 
Secondary instabilities are necessary to limit the growth of the runaway 
solutions and produce $z$-dependent states that represent steady turbulent
transport. 
However, Fig.~\ref{fig3} is a comparison between floating point and
double precision calculations for the mean squared velocity components 
indicating the sensitivity of the exponential solutions to random
perturbations generated by round-off noise and discretization errors. 

The simulations indicated that the breakdown of the exponentially growing
solutions for $w$ and $\Theta$ and the exponentially decaying solutions
for $u$ and $v$ is first signaled in these horizontal components.
It is not clear at this point what are the relevant scales involved in
this process for $Ra\gtrsim Ra_{c}$, although we observe that the growth of the
horizontal components destabilizes the exponentially growing modes,
leading to a fast redistribution of energy at all scales. 
At the peak, when $w$ and $\Theta$ reach their maxima, almost flat energy 
spectra are produced irrespective of the resolution adopted. 
Subsequently the high wave-vector dissipation takes over and the process is repeated, 
repeating the previous exponential growth.  
In summary, our simulations suggest that the exponential solutions may
in fact be attracting for a broad class of initial conditions, although 
subject to some finite amplitude instabilities.

There are other problems where extreme limits of models introduce
runaway solutions such as those appearing here. 
A distinct example is zero-Prandtl number Rayleigh-B\'{e}nard convection
\cite{EAS2,Thual} where the singular limit of the Boussinesq equations
admits exponentially growing solutions even in the presence of rigid
(albeit free-slip) boundaries. 
It was long ago observed in double diffusive convection in the 
absence of rigid boundaries that linearly unstable modes 
representing `salt fingers' are exact solutions of the
nonlinear equations \cite{Stern,Turner}.  

Another model implemented with fully periodic conditions to avoid
boundary layers, in which a similar nonlinear separation of variables
appears, is shear-driven turbulence.
In that model the fluctuations about an imposed
mean shear flow $\mbox{{\bf i}} S y$ obey
\begin{equation}
\bu_{t} + \bu\cdot\nabla\bu + S\,y\,\partial_{x} \bu +
\mbox{{\bf i}}\, S\, v + \nabla p = \nu \Delta\bu .
\label{shear}
\end{equation}
Fully periodic conditions cannot be implemented directly here, 
though, due to the presence of the incompatible operator  
$Sy\partial_{x}$ (explicitely non-periodic in $y$).
It was noted \cite{Rog} that periodic conditions {\it can} be imposed on
independent variables $x' = x-Syt$, $y'=y$, $z'=z$ and $t'=t$, and this
transformation has been used to perform numerical simulations of
``homogeneous shear flow'' \cite{PS}.

These periodic conditions also allow for an exact nonlinear separation 
of the cross-stream and stream-wise components.
Indeed, the change of variables implies 
\begin{equation}
\nabla \rightarrow \nabla' - \mbox{{\bf j}} S t \partial_{x'}
\quad \mbox{and} \quad 
\bu_{t} + Sy\partial_{x} \bu = \bu_{t'}.
\end{equation} 
Thus, acting on functions {\it only} of $y'$, $z'$ and $t'$, the operators
$\nabla = \nabla' = \mbox{{\bf j}} \partial_{y'} + \mbox{{\bf k}} 
\partial_{z'}$ and $\Delta = \Delta' = \partial_{y'}^{2} +
\partial_{z'}^{2}$.
Hence the system (\ref{shear}) separates for solutions 
depending only on $y'$, $z'$ and $t'$:
the  two-dimensional divergence-free velocity fields 
$\bv = \mbox{{\bf j}} v(y',z',t') + \mbox{{\bf k}} w(y',z',t')$
and pressure $p(y',z',t')$ satisfy the unforced Navier-Stokes equation
\begin{equation}
\bv_{t} + \bv \cdot \nabla' \bv + \nabla' p = \nu \Delta' \bv,
\label{2dns}
\end{equation} 
and the stream-wise component $u(y',z',t')$ evolves according
to the linear inhomogeneous equation
\begin{equation}
u_{t} + \bv \cdot \nabla' u + S v = \nu \Delta' u.
\end{equation} 
These equations do not (apparently) support unbounded exponentially
growing fields but they do display non-normal transient growth among their
fully nonlinear exact solutions.
Indeed, the decaying solutions of (\ref{2dns})
\begin{equation}
\bv = \Omega \, e^{-\nu (k_{2}^{2}+k_{3}^{2}) t'}
\, \sin(k_{2}y' + k_{3}z'+\phi)
\, [\mbox{{\bf j}} k_{3} - \mbox{{\bf k}} k_{2}]
\end{equation} 
produce a stream-wise flow of the form
\begin{equation}
u = (U - \Omega S k_{2} t') \, e^{-\nu (k_{2}^{2}+k_{3}^{2}) t'}
\, \sin(k_{2}y' + k_{3}z'+\phi)
\end{equation} 
where $U$ and $\Omega$ are set by initial conditions.
The peak amplitude $u_{peak} =  -e^{-1}\Omega S k_{3}/\nu(k_{2}^{2}+k_{3}^{2})$
(when $U=0$) may be extremely large at high Reynolds number when
$Sk_{3}/\nu(k_{2}^{2}+k_{3}^{2}) \gg 1$.
This is consistent with the behavior reported by Pumir and Shraiman based on 
their direct numerical simulations \cite{PS}:
``The transient regime is characterized by a violent growth of the kinetic energy \dots
\ While this growth eventually stops \dots \ the turbulent regime that follows exhibits
large fluctuations of spatially averaged quantities \dots \ Because of the unusually large
level of fluctuations, very long runs are necessary to get steady averages,
which explains why we chose to work at moderate resolution."

We conclude that the imposition of periodic boundary conditions may
admit an exact nonlinear separability that allows for larger fluctuations
than are possible in the presence of rigid boundaries. 
In full three-dimensional simulations of some of these systems, 
secondary instabilities are the only limiting processes that can lead
to finite statistically steady turbulent transport.
Among the delicate---and currently unresolved---issues is the question
of how sensitive the statistics of high $Ra$ (or $Re$) simulations
may be to numerical discretization and noise.
Nevertheless the HRB model,
although physically un-realizable because of the boundary conditions, 
has stimulated new and interesting experiments \cite{EXP}
where the effect of the thermal boundaries has been reduced to reveal
the $Nu\sim Ra^{1/2}$ scaling observed in the high Rayleigh number
simulations \cite{LT1,LT2}. 

\textit{Acknowledgments}: We thank B.~Castaing for useful
discussions and for sharing their results prior to publication.
E.C.\ thanks A.~Sevilla for many useful discussions. 
This work is part of the research program of FOM, which is supported by 
NWO and by the European Union (EU) under contract HPRN-CT-2000-00162. 
C.R.D. was supported in part by NSF-PHY0244859 and an 
Alexander von Humboldt Research Award.  
A.T. has been supported by an ORS award. 
C.R.D., J.D.G. and A.T. acknowledge the hospitality of the 2005 Program in
Geophysical Fluid Dynamics at Woods Hole Oceanographic Institution
where part of this work was completed.


\end{document}